# Genetic Influences on Brain Gene Expression in Rats Selected for Tameness and Aggression

Henrike O. Heyne[*,§], Susann Lautenschläger[§], Ronald Nelson[†], François Besnier[†,‡], Maxime Rotival[**], Alexander Cagan[*], Rimma Kozhemyakina[§§], Irina Z. Plyusnina[§§], Lyudmila Trut[§§], Örjan Carlborg[†], Enrico Petretto[**], Leonid Kruglyak[††,‡‡,***], Svante Pääbo[*], Torsten Schöneberg[§], Frank W. Albert[*,††]

[*] Department of Evolutionary Genetics, Max Planck Institute for Evolutionary Anthropology, Deutscher Platz 6, 04103 Leipzig, Germany

[§] Institute for Biochemistry, University of Leipzig, Medical Faculty, Johannisallee 30, 04103 Leipzig

[†] Swedish University of Agricultural Sciences, Department of Clinical Sciences, Division of Computational Genetics, Mailing address: Box 7078 SE-75007 Uppsala Sweden

[‡] Section of Population Genetics and Ecology, Institute of Marine Research, Bergen, Norway, Havforskningsinstituttet, Postboks 1870, Nordnes 5817, Bergen Norway

[**] MRC Clinical Sciences Centre, Faculty of Medicine, Imperial College London, Du Cane Road, London W12 0NN, UK

[§§] Institute of Cytology and Genetics, Siberian Branch of the Russian Academy of Sciences, 630090 Novosibirsk, Russia

[††] Department of Human Genetics, University of California, Los Angeles, Gonda Center, 695 Charles E. Young Drive South, Los Angeles, CA 90095, USA

[‡‡] Department of Biological Chemistry, University of California, Los Angeles, Gonda Center, 695 Charles E. Young Drive South, Los Angeles, CA 90095, USA

[***] Howard Hughes Medical Institute, University of California, Los Angeles, Gonda Center, 695 Charles E. Young Drive South, Los Angeles, CA 90095, USA

deceased:
I.Z.P. Irina Z. Plyusnina

.

# ABSTRACT

Inter-individual differences in many behaviors are partly due to genetic differences, but the identification of the genes and variants that influence behavior remains challenging. Here, we studied an F2 intercross of two outbred lines of rats selected for tame and aggressive behavior towards humans for more than 64 generations. By using a mapping approach that is able to identify genetic loci segregating within the lines, we identified four times more loci influencing tameness and aggression than by an approach that assumes fixation of causative alleles, suggesting that many causative loci were not driven to fixation by the selection. We used RNA sequencing in 150 F2 animals to identify hundreds of loci that influence brain gene expression. Several of these loci co-localize with tameness loci and may reflect the same genetic variants. Through analyses of correlations between allele effects on behavior and gene expression, differential expression between the tame and aggressive rat selection lines, and correlations between gene expression and tameness in F2 animals, we identify the genes *Gltscr2*, *Lgi4, Zfp40* and *Slc17a7* as candidate contributors to the strikingly different behavior of the tame and aggressive animals.

# INTRODUCTION

Behavioral differences among members of a species are in part due to genetic variation. The identification of the genes and variants that influence behavior remains challenging. In human genome-wide association studies of psychiatric and cognitive traits, the identified loci typically explain only a small fraction of the heritability, i.e. the additive genetic contribution to the trait (WATANABE *et al.* 2007; HOVATTA and BARLOW 2008; DEARY *et al.* 2009; OTOWA *et al.* 2009; CALBOLI *et al.* 2010; TERRACCIANO *et al.* 2010; FLINT and MUNAFO 2013; RIETVELD *et al.* 2013; SOKOLOWSKA and HOVATTA 2013). In experimental crosses in model species, especially between inbred lines, the identified loci (termed quantitative trait loci, QTL), often explain much more of the heritability (LYNCH and WALSH 1998). However, in this design the spatial resolution is limited – the QTL are wide and contain many, sometimes hundreds of genes. With the exception of a handful of identified genes with quantitative effects on behavioral traits (YALCIN *et al.* 2004; WATANABE *et al.* 2007; MCGRATH *et al.* 2009; BENDESKY *et al.* 2012; GOODSON *et al.* 2012), gene identification is particularly difficult for behavioral QTL that often have modest effect sizes (FLINT 2003; WILLIS-OWEN and FLINT 2006; WRIGHT *et al.* 2006; HOVATTA and BARLOW 2008).

The causal variants within a QTL can alter the protein sequence encoded by a gene or affect gene expression. Such regulatory variants can be identified as "expression QTL" (eQTL). An attractive approach for nominating candidate genes underlying QTL is therefore to identify genes within the QTL regions whose expression is influenced by an eQTL in a tissue of relevance to the phenotype (HUBNER *et al.* 2005; PETRETTO *et al.* 2006b; MACKAY *et al.* 2009). This approach has been successfully applied to quantitative traits in different species, e.g. yeast (PERLSTEIN *et al.* 2007), rats (HUBNER *et al.* 2005; HEINIG *et al.* 2010), mice (MEHRABIAN *et al.* 2005; SCHADT *et al.* 2005) and humans (MOFFATT *et al.* 2007; MUSUNURU *et al.* 2010). A few eQTL studies have also recently been carried out to identify candidate genes for behavioral QTL (HITZEMANN *et al.* 2004; DE JONG *et al.* 2011; SABA *et al.* 2011; KELLY *et al.* 2012).

Here, we studied two lines of rats (*Rattus norvegicus*) that, starting from one wild population, have been selected for increased tameness and increased aggression towards humans, respectively. These experimental populations derive from the work of Dmitry K. Belyaev who, along with his famous related work in foxes (BELYAEV 1979; TRUT 1999), initiated the rat lines to study the evolution and underlying biological mechanisms of tameness as a crucial first step in animal domestication (DIAMOND 1999; TRUT 1999; WIENER and WILKINSON 2011). For more than 64 generations, the tame rats were selected exclusively based on whether they tolerate or even seek out human contact, while the aggressive rats were selected to vigorously defend themselves from any attempted handling. This strong selection has resulted in two lines that respond extremely differently to attempts of handling. Whereas the aggressive rats attack and try to escape, the tame rats lack fear or aggression. They tolerate handling and sometimes actively explore the approaching hand. The strong behavioral selection has resulted in a number of differences in morphology, neurotransmitters and hormone levels (NAUMENKO et al. 1989; PLYUSNINA and OSKINA 1997; ALBERT et al. 2008).

Cross fostering experiments excluded postnatal maternal effects as a major influence on tameness and aggression in these animals, suggesting a substantial genetic contribution to the behavior (ALBERT et al. 2008; PLYUSNINA et al. 2009). To gain insight into the genetic basis of tameness and associated phenotypes, we earlier bred an F2 intercross between the tame and aggressive rat lines (ALBERT et al. 2009). Genetic mapping in 700 rats from this population revealed two genome-wide significant QTL for tameness. Three additional loci were found as part of an epistatic network influencing tameness (ALBERT et al. 2009). However, these QTL harbor hundreds of genes and individual genes influencing the behavior remain unknown.

The tame and the aggressive selection lines originate from one wild population. Throughout selection, the populations were kept as outbred as possible by avoiding matings between close relatives. Together with the limited number of generations and the modest population sizes since initiation of the lines, these characteristics make it unlikely that tameness and aggression are mainly caused by new mutations arising after the animals were caught. Instead, most genetic variants that were selected are expected to be due to standing variation already present in the wild population. We earlier confirmed the existence of considerable DNA sequence variation both between and within each of the two lines (ALBERT et al. 2011). Therefore, it seems plausible that alleles influencing behavior segregate not only between but also within the lines. QTL analyses that assume fixation of causative alleles could miss important sources of genetic variation.

Here, we used mRNA sequencing to quantify gene expression in the brains of 150 rats from the F2 intercross population. This data was used to map eQTL influencing differences in expression levels, using a QTL mapping method (Flexible Intercross Analysis, FIA (RÖNNEGÅRD et al. 2008)) that can account for segregating variation within the founders of a cross between two outbred lines. Using FIA, we also found four times as many loci influencing tameness than previously seen (ALBERT et al. 2009). We identify multiple genes within the tameness QTL whose expression in the brain is influenced by the genetic variants from the tame- and aggressive lines. By combining correlations of allele effects on behavior and gene expression, differential brain gene expression analyses between the tame and aggressive rat lines and correlation with tameness in F2 animals, we prioritize genes whose altered expression may be influencing tameness.

# MATERIALS AND METHODS

**Rat populations.** The tame and aggressive rat populations are derived from a long-term selective breeding experiment that was initiated in 1972 in the Institute of Cytology and Genetics in Novosibirsk, Russia (NAUMENKO *et al.* 1989). Two rat (*Rattus norvegicus*) lines were selected from one wild-caught population for the single criterion of increased tameness and increased aggression when confronted with a human hand. The original population consisted of about 200 animals. At each generation, the 30% most tame and most aggressive individuals were selected for further breeding. Inbreeding was avoided in both strains by avoiding mating between close relatives. In 2005, subpopulations from both lines were transferred to Leipzig to study the genetic basis of the tameness (ALBERT *et al.* 2008; ALBERT *et al.* 2009; ALBERT *et al.* 2011; ALBERT *et al.* 2012). During all care-taking procedures and experiments, animals from different lines and generations were treated identically. The study was approved by the regional government of Saxony (TVV Nr. 29/95).

**Selection of F2 animals based on extreme tameness/aggression.** To perform genetic mapping of loci influencing tameness and aggression, we have earlier generated an F2 intercross population of ~700 rats from a cross between the tame and the aggressive rats (ALBERT *et al.* 2009). Briefly, in that study F2 rats were assigned a "tameness score" based on a behavioral test (ALBERT *et al.* 2008) where rats were exposed to a standardized handling procedure, and their responses videotaped and scored. To summarize a rat's behavior in this test, principal component analysis (PCA) was performed on a set of 11 individual measures. The first principal component provides a quantitative measurement for tameness (ALBERT *et al.* 2009). For the present study, we used stored organs from the 75 most tame and the 75 most aggressive F2 animals.

**Tissue collection.** Animals were earlier anaesthetized by $CO_2$ and killed by cervical dislocation. (ALBERT *et al.* 2009) Tissues were collected, snap-frozen in liquid nitrogen and stored at -80°C. Before freezing, each brain was dissected to separate the brainstem and cerebellum from the fore- and midbrain.

For the current study, brains were cut with a scalpel along the mid-sagittal plane in two hemispheres at -20°C. Before dissection tissues were left to equilibrate to -20°C for at least half an hour to obtain the best consistency during the cutting. To protect RNA quality, brains were never completely thawed during dissection. Complete right brain hemispheres comprising tel-, di- and mesencephalon were used. Frozen hemispheres were homogenized in 5 ml of TRIzol (Invitrogen, Darmstadt, Germany) in glass tubes using a Schütt Homogenizer with 3000 rpm for 13 rounds. 2.5 ml TRIzol was added afterwards to obtain a final volume of 7.5 ml TRIzol per brain hemisphere.

**cDNA library preparation.** RNA was extracted by chloroform extraction and purified with the Qiagen RNEasy MinElute Clean-up kit according to the manufacturer's instructions. All RNA samples were of high quality as judged by Agilent Bioanalyzer (Agilent Technologies, Böblingen, Germany). "RNA Integrity Numbers" (RIN) (SCHROEDER *et al.* 2006) for all samples ranged from 8.5 to 10, where 10 corresponds to maximum RNA quality (Supplementary Table S1). RNA-seq libraries were generated according to in-house protocols, details see (KUHLWILM *et al.* 2013). mRNA was extracted from 10 μg of total RNA by capture on poly-T covered magnetic beads, chemically fragmented and used as template for cDNA synthesis using random hexamer primers. Double-stranded cDNA was blunt-ended and Illumina sequencing adapters were ligated to the cDNA. To minimize potential batch effects, balanced numbers from the tame and the aggressive groups of F2 animals were processed together throughout RNA extraction and library generation and distributed evenly across sequencing runs.

**Sequencing.** Libraries were adjusted to a concentration of 10 nM and individuals distributed into three different pools, each consisting of equal numbers of tame and aggressive individuals. They were then sequenced to a median coverage of 10.8 million reads per animal (standard deviation = ± 0.8 million reads per animal) on Illumina Genome Analyzer 2 machines (Illumina, Inc., San Diego, USA) with 76 bp single end reads. We called bases using Ibis (KIRCHER *et al.* 2009). Adapter sequences were trimmed, adapter dimers excluded and the reads filtered for quality. All RNA-Seq read data will be made publicly available upon acceptance of the paper.

**Read mapping and gene expression quantification.** Reads were mapped to the rat genome (assembly and gene annotation RGSC 3.4) with the programme TopHat, version 1.4.1 (TRAPNELL *et al.* 2009). We ran TopHat with the --transcriptome index option and provided the program with a transcriptome sequence file obtained from Ensembl release 69. With this option, TopHat first aligns reads to the virtual transcriptome before mapping unmapped reads to the whole genome.

Gene expression was then quantified using the program Cufflinks (TRAPNELL *et al.* 2010) version 1.3.0, with the following parameters: We employed the fragment bias correction algorithm, which aims to account for sequence specific biases introduced by the use of random hexamer primers (ROBERTS *et al.* 2011). We further provided Cufflinks with the median read length for each library as measured with the Agilent BioAnalyzer before sequencing. Further, we supplied Cufflinks with a reference annotation file to quantify whole gene and different isoform expression, using the Ensembl release 69 gene annotation. Any alignments not compatible with any reference transcripts were ignored. We thus did not search for novel genes, and did not attempt to alter the gene models represented in the Ensembl annotation. Unless otherwise noted, we used FPKM expression values (TRAPNELL *et al.* 2010) for further analyses. In FPKM, the RNASeq data is normalized to the length of each gene and the total number of aligned reads in the library (TRAPNELL *et al.* 2010). FPKM values were transformed using log2 (FPKM+1). All further analyses were done in the R environment .

**Gene expression pre-processing.** We retained for analyses all genes for which Cufflinks reported expression levels, resulting in quantifications for 24,139 genes. To avoid confounding by potential batch effects (LEEK *et al.* 2010), we corrected the expression data for batches defined by groups of samples processed on the same date (Supplementary Table S1). This was done by fitting a linear model with gene expression as the response variable and batch as the predictor variable. We used the residuals of the model as batch-corrected gene expression traits. To additionally account for unknown unwanted sources of variation, we computed surrogate variables of the gene expression matrix by using the R package sva (LEEK and STOREY 2007; LEEK *et al.* 2012). We used the two surrogate variables reported by the software as covariates for expression QTL mapping. We used the phenotypes tameness and adrenal gland weight (ALBERT *et al.* 2009) as covariates when calculating the surrogate variables. To rule out the possibility that we might accidentally correct for true genetic regulators that might influence many gene expression traits, we also performed QTL mapping for the values of the two surrogate variables. The two SVA did not significantly map to any region of the genome. SVA-corrected expression data yielded more eQTL than either a correction for sex or a correction for sex and RNA library processing batch (Supplementary Table 2). We also used the animals' sex as a covariate during eQTL mapping.

**Genetic map construction.** Genotypes at 152 microsatellite and 49 single-nucleotide polymorphism (SNP) markers were used to build a 1728 cM genetic map as in (ALBERT *et al.* 2009). Genotypes, genetic markers, marker positions and pedigree data were the same as for the QTL mapping of tameness described previously in (ALBERT *et al.* 2009). The probability of each F2 allele to originate from either the tame or the aggressive line were re-computed using an updated algorithm where haplotypes and missing marker information were inferred iteratively in order to reconstruct any missing genotypes. First, haplotypes were inferred from genotype and pedigree data with a custom implementation of the method described in (BESNIER and CARLBORG 2009). In our implementation, haplotype reconstruction steps were alternating with missing genotype inference steps. The two steps were run iteratively for each pedigree generation, in turn, until no new haplotype or missing genotype could be inferred. The probability of each F2 allele to originate from either the tame or the aggressive line were then re-inferred from phased genotype and pedigree, using the Identity By Descent (IBD) matrix calculated as in (PONG-WONG *et al.* 2001) and implemented in (BESNIER and CARLBORG 2007).

**eQTL mapping using HKR.** We searched for eQTL on autosomes in steps of 1 cM using Haley Knott regression (HKR) (HALEY and KNOTT 1992) as implemented in the R package R/qtl (BROMAN *et al.* 2003). The outbred cross dataset was prepared for use in R/qtl using the R package qtl.outbred (NELSON *et al.* 2011). We used LOD (logarithm of odds) scores to quantify how strongly genetic variation across the genome influences the expression of a gene. To determine the significance threshold for QTL, we performed 1,000 permutations across all 14,000 genes with sufficiently abundant expression (see below) by randomizing genotypes with respect to phenotypes and performing full genome-scans for the permuted datasets. To calculate false discovery rates (FDRs), we divided the number of expected eQTL (as determined from the permutations) by the number of eQTL observed in the real data at a given LOD threshold (BENJAMINI and YEKUTIELI 2005). We found that a LOD threshold of 5.28 results in an FDR of 5% and used this value as the significance threshold for further analyses. We report confidence intervals for QTL location as 1.8 LOD drop intervals as recommended for F2 intercross designs in (BROMAN and SAUNAK 2009).

**Influence of sequencing depth on eQTL mapping.** To evaluate to what extent sequencing depth influences the number of detected eQTL, we randomly selected 5,000 genes, downsampled their expression data to different sequencing depths, and repeated the eQTL mapping in this trimmed dataset. At each depth, we randomly permuted genotypes relative to phenotypes 10 times across the 5,000 genes and determined the 99.9% quantile of the resulting distribution of LOD scores. This LOD score was used as the significance threshold for eQTL detection at the given depth. We calculated false discovery rates and recorded the number of eQTL that would have been found had we sequenced to this depth of coverage using the same procedure as described above (Supplementary Figure S1).

**False discovery rates as a function of gene expression level.** To assess at which expression level genes can be measured accurately enough to give robust eQTL mapping results we calculated false discovery rates in bins of increasing gene expression levels (Supplementary Figure S2). Genes were grouped according to their expression level into 24 bins each containing 1,000 genes, except for the bin with the lowest expression which contained 1,139 genes. We performed 10 permutations within each of the 24 bins, determined the 99.9% quantile of permuted LOD scores and used this LOD score as the eQTL detection threshold for the given bin. We found that the FDR stayed consistently below 5% among the 14,000 most highly expressed genes (Supplementary Figure S2). At a 5% FDR, we identified 3.2 eQTL per 100 genes when considering the 14,000 highest expressed genes as opposed to 1.7 eQTL per 100 genes when considering all genes. For all later analyses, we only used the 14,000 most highly expressed genes.

**Flexible Intercross Analysis.** "Flexible Intercross Analysis" (FIA) (RÖNNEGÅRD *et al.* 2008) is a QTL analysis method for analyses of data from intercross populations that does not assume that the founder lines of are fixed for alternative QTL alleles. Here, we used FIA, as implemented in R scripts kindly provided by L. Rönnegård, to perform genome-scans for tameness and gene expression. For tameness, significance thresholds were determined by 1,000 random permutations. For expression data, hundreds of permutations for all genes were not computationally feasible even on a computer with hundreds of nodes. Instead, we approximated the score distribution for QTL under the null-hypothesis using information from 30 permutations for each of 2,000 randomly sampled genes. False discovery rates were estimated as described above for HKR.

**FIA QTL confidence intervals.** We defined FIA eQTL confidence intervals by using information from eQTL that were detected by both FIA and HKR. We used the approx.fun function in R to fit a function relating the FIA mapping score to the width of the HKR confidence interval. Using this function, we interpolated confidence intervals for all FIA QTL. To avoid unrealistically small confidence intervals for the most highly significant eQTL we set the minimum size of the confidence intervals to 10 cM (Supplementary Figure S3).

eQTL were treated as local when the gene's physical position overlapped their confidence interval. Genetic positions in cM were interpolated to the physical Mb scale and vice versa with the approx.fun function using coordinates given in the Supplementary Table S2 of (ALBERT *et al.* 2009) that contains the genetic and physical position of the markers that were used to build the genetic map.

eQTL and tameness QTL were treated as overlapping when their confidence intervals overlapped by at least one cM. In the test of whether windows within tameness QTL had more eQTL than the rest of the genome we used non-overlapping sliding windows with a width of 18 cM, which is the median confidence interval size of HKR eQTL.

**Regional clustering of eQTL.** We counted the number of genes with local eQTL in a sliding window analysis. Windows were advanced in 1 cM steps and had a width of 18 cM, which was the median confidence interval for eQTL in the HKR analysis. When calculating correlations between the numbers of genes located in an interval with the number of local eQTL, we used non-overlapping sliding windows of 18 cM to minimize dependencies between adjacent data points.

To test whether there are positions with more distant eQTL than expected, we determined the number of distant eQTL that would be expected to map to a given position in the genome if eQTL were positioned at random. We generated 1,000 sets of random genomic positions for the distant eQTL while keeping the number of eQTL per chromosome unchanged, (Supplementary Figure S4). We used the 99% quantile of the resulting distribution as the threshold to call significant eQTL clusters.

**Co-expression networks.** WGCNA (LANGFELDER and HORVATH 2008) was used on the set of 14,000 expressed genes to detect clusters. Biweight correlation was used to identify clusters robust to outliers and all other parameters were left to default. GO enrichment was performed using DAVID (HUANG et al. 2008). Transcription factor binding site enrichment was performed using PASTAA (ROIDER et al. 2009): Sequences of 2x500bp regions around the transcription start site of each gene were extracted from the BN genome and scored for JASPAR and transfac public transcription factor matrices. Clusters were then tested for enrichment of high binding affinity values.

**Correlating allele effects of tameness QTL and expression QTL.** We performed two scans for allele effect correlations. In both scans, both local and distant eQTL were used. In the first scan, we tested for correlation between those eQTL that overlap one of the eight FIA tameness QTL. The correlations were determined at the peak positions of the tameness QTL, irrespective of the location of the respective eQTL peaks. In the second scan, we tested all eQTL, irrespective of whether they overlapped a tameness QTL. These correlations were determined at the peaks positions of the eQTL. The two scans thus differ in several respects: the set of eQTL in the analyses, the precise positions tested (these can differ between the two scans even for pairs of tameness and eQTL represented in both scans), and the corrected significance threshold (due to different numbers of tests).

**Differential gene expression in brains of animals from the tame and aggressive founder lines.**
Right brain hemispheres were harvested from five adult male rats from the tame and from the aggressive selection lines, respectively. mRNA was extracted as described above and fluorescently labeled and hybridized to Affymetrix microarrays (GeneChip Rat Genome 230 2.0) following the manufacturer's instructions. We mapped array probes to genes using re-annotated chip definition files (DAI *et al.* 2005). Array normalization was performed using robust multiarray average (RMA) (IRIZARRY *et al.* 2003) and genes called as expressed using the Affymetrix MAS5 algorithm. Both RMA and MAS5 were used as implemented in the R bioconductor 'affy' package (BOLSTAD *et al.* 2003). Only genes where at least 4 of the 5 animals in a group showed expression according to MAS5 were analyzed. We tested for differential expression between the tame and aggressive animals using T-tests. The FDR was determined by calculating all permutations that are possible in two groups of five samples. For a range of p-value cutoffs, we calculated the expected number of false positive tests as the median number of tests exceeding a given cutoff across permutations. The FDR was calculated as this expected number of false positive tests divided by the number of positive tests observed in the real data.

**Correlation of gene expression levels and tameness in F2 animals.** We tested for correlation (method: Spearman – rank correlation) between gene expression levels and tameness. The FDR was determined by comparing the observed data to 100 permuted data sets. We calculated the FDR separately for each tameness phenotype.

# RESULTS

**RNA sequencing.** We selected the 75 most tame and the 75 most aggressive animals the 700 F2 intercross rats bred for our earlier study (ALBERT *et al.* 2009) for RNA sequencing. Non-stranded Illumina sequencing libraries were generated from polyA-enriched RNA from cerebral hemispheres that were frozen as part of our earlier characterization of these animals. The RNA-Seq libraries were sequenced to a median coverage of 10.8 million reads per animal (standard deviation (sd) = 0.8 million reads). Of these, an average of 9.0 million reads (83% of total reads) mapped to the rat genome (sd = 0.6 million reads). Out of these, an average of 61.7% (sd = 2.1%) mapped unambiguously to known exons. Down-sampling analyses were used to evaluate the power of the study, and although these suggest that more eQTL could be found by increasing sequencing coverage (Supplementary Figure S1), the available coverage is expected to be sufficient to discover 100s of eQTL. Our strategy of selecting F2 animals with extreme behaviors did not lead to notably skewed or bimodal gene expression distributions (Supplementary Figure S5), suggesting that this strategy does not complicate eQTL mapping.

**Fixed genetic differences between the tame and aggressive rat populations influence brain gene expression levels.** We began our search for eQTL using a method that assumes that the two founder populations are fixed for QTL alleles with different effects on expression (Haley Knott regression (HKR) (HALEY and KNOTT 1992); this method was also used in our earlier mapping study (ALBERT *et al.* 2009)). Because the accuracy of RNA-Seq increases with increasing gene expression abundance, we limited our search to genes where eQTL can be detected with high statistical power (see Methods). In total, 443 eQTL were detected at a False Discovery Rate (FDR) of 5% using HKR (Supplementary Table S3). Thus, genetic differences between the tame and the aggressive rats result in mRNA expression differences at hundreds of genes.

**Genetic variation within the selection lines influences tameness and brain gene expression levels.** In order to explore the contribution of within-line genetic variation on behavior and gene-expression, we used "Flexible Intercross Analysis" (FIA) (RÖNNEGÅRD *et al.* 2008), a QTL mapping method that is able to detect loci that are not fixed between the tame and aggressive founders.

We first applied FIA to tameness and aggression in all 700 F2 animals and found 8 QTL affecting tameness and aggression at a 2% permutation-based FDR (Figure 1, Table 1). These QTL include the two previously identified QTL (ALBERT *et al.* 2009). At the most significant FIA tameness QTL on chromosome 1 (corresponding to the "Tame-1" locus identified in (ALBERT et al. 2009)), the alleles carried by the tame and the aggressive individuals used to found the F2 pedigree have clearly different effects on tameness (Figure 1). However, more than one "aggressive" allele appears to exist among the four aggressive founder animals (ALBERT et al. 2011), and one of the four tame founders appears to be heterozygous for one "tame" and one "aggressive" allele. Nevertheless, the overall allelic differentiation between the tame and the aggressive founder individuals explains why this locus was also the strongest QTL identified by HKR. By contrast, the FIA QTL at the end of chromosome 10 that had the third highest significance did not correspond to a HKR QTL (Figure 1). The allelic effects at this locus are highly heterogeneous, with at least one tame and one aggressive allele segregating in both the tame and the aggressive founder animals. Thus, allowing causative alleles to segregate within the lines quadruples the number of detected tameness QTL, suggesting that the genetic basis of this behavior is considerably more complex than previously appreciated.

Next, we asked how segregating allelic variation affects variation in brain gene expression levels. When using FIA for eQTL mapping, we detected 689 FIA eQTL influencing the expression of 670 genes at an FDR of 5% (Figure 2, Supplementary Figure S6, Supplementary Table S4). Of these, 411 were also detected by HKR. Conversely, of the 443 HKR eQTL, all but 32 were detected by FIA (Figure 3). These 32 eQTL had low LOD scores (LOD = "logarithm of the odds") close to the genome-wide significance threshold, lower than the HKR eQTL identified with both HKR and FIA (Supplementary Figure S7A, Wilcoxon-Rank test: $p = 6.9e-14$). To be conservative, we excluded the 32 eQTL identified only by HKR from further analyses. While eQTL that were found only by FIA had, on average, lower scores than eQTL found with both HKR and FIA (Wilcoxon-Rank test $p < 2.2e-16$), many of them had FIA mapping scores that were highly significant (Supplementary Figure S7B). Thus, FIA increased the number of detected eQTL by 68%, including some loci with strong effects on gene expression that were missed by HKR because their alleles segregate within the selection lines.

**Genomic architecture of mRNA expression.** Of the 689 FIA eQTL (Supplementary Table S4), 561 were local eQTL close to the genes they influenced and 128 were distant eQTL that did not overlap the position of the affected gene. Local eQTL had higher mapping scores than distant loci (Wilcoxon-Rank test, $p = 1.7e-07$). This is consistent with previous work showing that while both local and distant eQTL exist, local eQTL typically have stronger effects (PETRETTO et al. 2006a; MACKAY et al. 2009; VAN NAS et al. 2010).

Clusters of *trans*-eQTL can be indicative of networks of co-regulated genes under common genetic control by a single locus ((HUBNER *et al.* 2005; HEINIG *et al.* 2010; LANGLEY *et al.* 2013)). We therefore explored whether the distant eQTL are enriched in certain regions of the genome. One region contained more distant eQTL than expected if the eQTL were distributed randomly throughout the genome. This excess was, however, small (7 eQTL per window, ≤ 6 expected by chance, Supplementary Figure S4A). Further, the region did not overlap with any tameness QTL (Supplementary Figure S4B). To directly investigate co-expression networks and their genetic regulation, we applied Weighted Gene Co-expression Network Analysis (WGCNA) (LANGFELDER and HORVATH 2008). The method grouped 12,616 of the 14,000 genes in the analyses into 54 co-expression clusters (Supplementary Table S5; Supplementary Note). For each cluster, we computed the first principal components of the expression levels of the genes in the cluster and asked how they might relate to tameness. The expression of five clusters was correlated with tameness across the 150 F2 animals (Supplementary Table S5). Further, mapping the expression of the 54 clusters using FIA revealed QTL for several of the clusters, but none of these QTL overlapped with a tameness QTL.

**Overlap of QTL affecting tameness and gene expression in the brain.** Genes whose expression levels in the brain are influenced by local eQTL that overlap tameness QTL are candidates for contributing to the differences in behavior. Of the 689 FIA eQTL, 212 (influencing the expression of 207 genes) overlapped with a tameness QTL (Figure 4). The genomic region spanned by *Tame-1* (the most significant tameness QTL) locus contained more eQTL than other regions of similar size in the rest of the genome (Wilcoxon-Rank test, p = 0.007). This observation is a reflection of the fact that *Tame-1* is located in the most gene-dense region of the linkage map (Supplementary Figure S8).

Overall, the scores of those eQTL that overlapped a tameness QTL were not different from those that did not (Wilcoxon-Rank test, p-value = 0.4; Figure 4). However, the *Tame-1* region was enriched for highly significant eQTL (Figures 4 and 5). For example, 11 of the 79 eQTL (14%) with a score of at least 1,000 were located in *Tame-1* although this locus contains only 3% of the 14,000 genes in the analyses. Thus, while many eQTL overlapped a tameness QTL, the strongest tameness QTL *Tame-1* stands out in that it spans the region with the highest gene density in the genome (Supplementary Figure S8) and also contains many strong local eQTL.

**Allele effect correlations between tameness QTL and eQTL.** An overlap of QTL for tameness and gene expression does not imply that the two QTL are necessarily caused by the same sequence variants. However, where there are multiple alleles at FIA QTL, a correlation between the allele effects of overlapping tameness QTL and eQTL is expected if they are driven by the same genetic variants. If the overlap is with a local eQTL, causative variation at the gene itself may underlie the tameness QTL. If the overlap is with a distant eQTL, the gene regulated in *trans* need not itself contain sequence variants that influence tameness, but its expression may be part of the biological pathways that shape this behavior.

We performed two scans for allele effect correlations. First, we conducted a targeted analysis of the FIA tameness QTL and the 212 FIA eQTL that overlap them. In this scan, the allele effects of 29 genes correlated significantly with the tameness allele effects (Benjamini Hochberg (BH) - adjusted p < 0.05, Figure 6A). Of these 29 correlations, 25 involved *Tame-1*, and three involved distant eQTL (Supplementary Table S4).

To explore if these allele effect correlations were specific to the significant tameness QTL, we conducted a second scan in which we tested all local eQTL for allele effect correlations with tameness, irrespective of the evidence for a tameness QTL at the respective position. Out of 23 significant correlations, 18 (78%) occurred at tameness QTL. Thirteen of these involved eQTL that overlapped *Tame-1*, and two additional correlations involved eQTL for genes situated very close to *Tame-1* (Figure 6B). Thus, correlations between allele effects on tameness and on gene expression mostly occur at loci with significant effects on the behavior.

**Correlations between gene expression levels and tameness.** Across the 150 F2 animals, we found four genes whose expression levels correlated significantly with tameness (correlation strength exceeds that seen in any of 100 permutations, Supplementary Figure S5, Supplementary Table S6). Of these, the expression of *Gltscr2* and *Zfp40* correlated positively with tameness. Both genes had eQTL at the *Tame-1* peak and showed allele effect correlations with *Tame-1* (Figure 7). The other two genes did not have a significant eQTL.

**Differential expression between the tame and aggressive selection lines.** We searched for differentially expressed genes between the tame and the aggressive rat lines using two datasets: a published RNA-Seq dataset of frontal cortex tissue [47] and a microarray-based comparison of cerebral hemispheres first reported here. We found 11 genes that were differentially expressed between the tame and aggressive founder animals and also had eQTL overlapping with tameness QTL (Supplementary Table S4). Five of these genes (*Dhdh*, *Slc17a7*, *Lgi4*, *Gltscr2* and *Dll3*) also showed significant allele effect correlations with a tameness QTL (Figure 7, Supplementary Table S7).

There was one gene (*Gltscr2*) that satisfied all the criteria we considered for nominating candidate genes for influencing tameness. In addition, several genes satisfied two or more criteria (Figure 7, Supplementary Table S7).

# DISCUSSION

We have explored the influence of genetic variation on brain gene expression levels in a cross between two outbred lines of rats that had been selected from a common base-population for strong tame and aggressive behavior towards humans. We further examined how this regulatory gene-expression variation relates to genetic influences on tameness and aggression and used the expression data to prioritize genes that may play a causal role in these behaviors.

We used two QTL mapping approaches that differ in their assumptions about the genetic homogeneity of the founder-lines. The Haley-Knott regression (HKR) assumes fixation for alternative causal alleles in the divergent founders. Using this method, we identified two tameness QTL (ALBERT *et al.* 2009) and hundreds of eQTL. Because the two selection lines are outbred and display considerable within-line genetic variation (ALBERT *et al.* 2011), we also used a QTL mapping method that does not require causative alleles to be fixed between the tame and aggressive founders (RÖNNEGÅRD *et al.* 2008). Using this method, many additional tameness QTL and eQTL were detected. These results indicate that the genetic architecture of behavior in this cross is complex and influenced by multiple QTL, a majority of which have alleles that still are segregating in the founders. Further, there are profound differences in the brain transcriptome that are under genetic control.

Rodent behavioral QTL studies typically identify multiple QTL with small effect sizes of less than 10% of the phenotypic variance of the trait (FLINT 2003; WILLIS-OWEN and FLINT 2006). For example, a QTL study in heterogenous stock mice derived from eight progenitor strains identified 205 loci contributing to anxiety (GOODSON *et al.* 2012). As a QTL needs to exceed a certain effect size in order to be detectable, the number of tameness QTL we have identified is an underestimate of the number of loci contributing to the phenotypic differences between the lines. Based on work on complex traits in other species, we might expect many more loci with effects too small to be detected in our current sample to influence a complex behavioral trait like tameness (e.g., (YANG *et al.* 2010; HUNT *et al.* 2013)). Further, selection experiments in multiple species (drosophila (BURKE *et al.* 2010), maize (LAURIE *et al.* 2004), chickens (JOHANSSON *et al.* 2010; PETTERSSON *et al.* 2013)) demonstrated that many loci typically respond to strong selection, as expected when trait variation is due to variants at many genes. It thus seems plausible that many genetic variants with small effects contribute to differences in several biological pathways which together shape complex behavioral traits such as tameness and defensive aggression. Epistatic interactions, where the effect of one locus depends on the genotype of another locus (PHILLIPS 2008) may also play a role, especially given that we have earlier detected epistatic interactions between tameness QTL (ALBERT *et al.* 2009). The large number of genes and pathways described in the literature influencing behavioral traits related to tameness such as anxiety and aggression makes it plausible that genes in multiple biological pathways could have reacted to the strong selection in these rats (HOVATTA and BARLOW 2008; LE-NICULESCU *et al.* 2011).

A major goal of our analyses was to compare genetic influences on brain gene expression to those on tameness and aggression. Genetic influences on brain gene expression are plausible contributors to behavioral phenotypes. We found numerous eQTL that overlapped the eight tameness QTL (Supplementary Table S4). In particular, the strongest tameness QTL (*Tame-1*) contained multiple highly significant eQTL with allele effects that were correlated with those on tameness. This result indicates that *Tame-1* might harbor several genes influencing behavior through altered gene-expression in the brain. QTL identified in F2 crosses frequently fractionate into loci of more minor effects during fine-mapping (DEMAREST *et al.* 2001; MOZHUI *et al.* 2008) and contain multiple causative variants (BAUD *et al.* 2013; FLISTER *et al.* 2013). An advanced intercross using large numbers of rats to generate more recombination events will be necessary to investigate whether variation at more than one gene underlies Tame-1 (DARVASI and SOLLER 1995).

We utilized the ability of FIA to infer the effects of multiple alleles at both tameness QTL and eQTL to prioritize potentially causative genes among genes with eQTL that overlap a tameness locus. The rationale is that if multiple haplotypes of a gene at an overlapping QTL/eQTL locus influence both tameness and the expression of a gene, the relative strengths of their allele effects on tameness and expression should be correlated. In contrast, if the QTL/eQTL locus is due to different variants affecting tameness and gene expression in the brain, it is less likely that their effects are correlated. Thus, identification of pairs of eQTL and tameness QTL with segregating alleles and correlated genetic effects allows us to select more likely causal genes (NICA *et al.* 2010). For example, of 43 eQTL that overlap *Tame-1*, only 10 had significant allele effect correlations in both our two scans, implying these 23% of the genes as more likely to be causal. This type of prioritization would not have been possible in a cross between two inbred lines.

In addition to correlation of allele effects on behavior and gene expression, we considered differential brain gene expression between the tame and aggressive rat lines and correlation of gene expression traits with tameness across F2 animals to identify the most plausible candidate causal genes. The genes highlighted by these analyses (Figure 7) include *Lgi4*, *Gltscr2*, *Zfp40* and *Slc17a7*, all of which are located in *Tame-1* and had significant eQTL allele effect correlations with this tameness QTL. With the exception of *Slc17a7*, none of these genes have to date been implicated in the regulation of a behavioral trait.

*Lgi4* is involved in glial cell proliferation and axon myelination in the peripheral nervous system (KEGEL *et al.* 2013). There is also evidence for a biological role in the central nervous system: polymorphisms in *Lgi4* have been associated with central epilepsy conditions (GU *et al.* 2004; ISHII *et al.* 2010). We found *Lgi4* to be differentially expressed between the tame and aggressive selection lines. Alleles from the tame line decreased *Lgi4* expression both in the contrast between the selected lines and in the F2 population.

*Gltscr2* encodes a regulator of the tumor suppressor genes *p53* (GOLOMB and OREN 2011) and *PTEN* (YIM *et al.* 2007) that shows lower expression in several tumors including gliomas (KIM *et al.* 2008). It was also described to have functions in embryogenesis and embryonic stem cell survival (SASAKI *et al.* 2011). Alleles from the tame line increased *Gltscr2* expression both when comparing the two selection lines and in the F2 population. In addition, *Gltscr2* expression was positively correlated with tameness across the F2 animals.

*Zfp40* (Zinc finger protein 40) is a little studied gene that it is expressed in murine embryonic nervous system development (NOCE *et al.* 1993). In the F2 animals, alleles from the tame line increased *Zfp40* expression and the expression of *Zfp40* was significantly and positively correlated with tameness.

*Slc17a7* belongs to a family of vesicular glutamate transporter genes and has been previously implicated in the regulation of behavior. Mice that carry a hemizygous knockout such that they express only half the amount of *Slc17a7* compared to wild type mice, are more anxious (TORDERA *et al.* 2007) and more vulnerable to chronic mild stress (GARCIA-GARCIA *et al.* 2009; FARLEY *et al.* 2012). Here, the expression of *Slc17a7* in brain hemispheres in rats from the tame selection line was twice that in the aggressive rats. However, the eQTL acted in the opposite direction: alleles from the aggressive founder animals increase *Slc17a7* expression. Further, *Slc17a7* expression does not correlate with tameness in the F2 animals, and *Slc17a7* is located at some distance (13 cM) from the peak of *Tame-1*. Thus, although it is a promising candidate gene, precisely how *Slc17a7* may contribute to the tameness and aggression remains unclear.

We found several genes that were not themselves located in tameness QTL, but whose expression was influenced by distant eQTL that did overlap tameness QTL. *Trans* acting variants in the tameness QTL might lead to differential expression of these genes, which in turn might shape the physiological pathways that regulate tame or aggressive behavior. An example of such a gene is *Cyp7b1*, whose expression is influenced by an eQTL that overlaps a tameness QTL on chromosome six. *Cyp7b1* encodes a 25-hydroxycholesterol 7-alpha-hydroxylase that is involved in DHEA metabolism (YAU *et al.* 2003). Interestingly, a key enzyme involved in the metabolism of this neurosteroid is located in a fixed duplication in 20 tame founder rats in *Tame-1* (Alexander Cagan, personal communication). It is interesting that genetic variation in two independent tameness QTL is connected to the DHEA system, making this system an interesting candidate for further studies into the regulation of tameness and aggression.

The gene *Htr3a* encodes a subunit of the ionotropic serotonin receptor type 3 (5-HT3). It has earlier been shown that there are differences in serotonin level between the tame and aggressive founder lines in different brain regions (NAUMENKO *et al.* 1989; ALBERT *et al.* 2008). Further, serotonin and its metabolite were higher in several brain regions in a line of foxes selected for tameness than in foxes not selected for behavior (POPOVA *et al.* 1976; POPOVA *et al.* 1991). 5-HT3 receptor density is also greater in hamsters showing impulsive aggression (CERVANTES and DELVILLE 2009) and *Htr3a* knockout mice exhibit changes in anxiety-like behavior and hypothalamus- pituitary- adrenal (HPA) axis response to stress (BHATNAGAR *et al.* 2004). In this study, *Htr3a* is differentially expressed between the tame and aggressive founders at a 10% FDR and weakly correlated (but not robust to multiple testing) with tameness in the F2. Also, the expression of the gene *Htr3a* is influenced by two distant eQTL, one of which overlaps with the second strongest FIA tameness QTL on chromosome 18. The allele effect correlations at this tameness QTL were nominally significant (but not robust to multiple testing). Although the evidence for *Htr3a* is marginal in each independent analysis, the coherency in the results indicates that it is a useful gene for further studies into the basis of tameness in this population.

In summary, we identified hundreds of genomic loci that influence gene expression levels in the brains of rats that had been bred for tame and aggressive behavior towards humans. Differences in brain gene expression are an attractive avenue by which genetic differences between and within these two rat lines may contribute to the strong differences in behavior. We also found six loci with effects on tameness that had been missed by earlier work, illustrating that the genetic architecture of behavior is complex in this artificially selected population. The expression of several genes inside the strongest tameness QTL was influenced by local eQTL with very strong effects. These genes are promising candidates for functional follow-up work to study if and how they contribute to tameness and aggression.

## ACKNOWLEDGEMENTS

We are grateful to Martin Kircher, Martin Kuhlwilm, Michael Dannemann, Rigo Schultz, Udo Stenzel and other past and present members of the Max Planck Institute for Evolutionary Anthropology for help with bioinformatic analyses. This work was supported by the Max Planck Society, an Sfb 1052 (DFG) and IFB AdiposityDiseases Leipzig (BMBF) to T.S., a DFG research fellowship AL 1525/1-1 to F.W.A., a Studienstiftung des deutschen Volkes stipend to H.O.H., MD1 and MD2 (IFB AdiposityDiseases Leipzig) stipends to H.O.H.

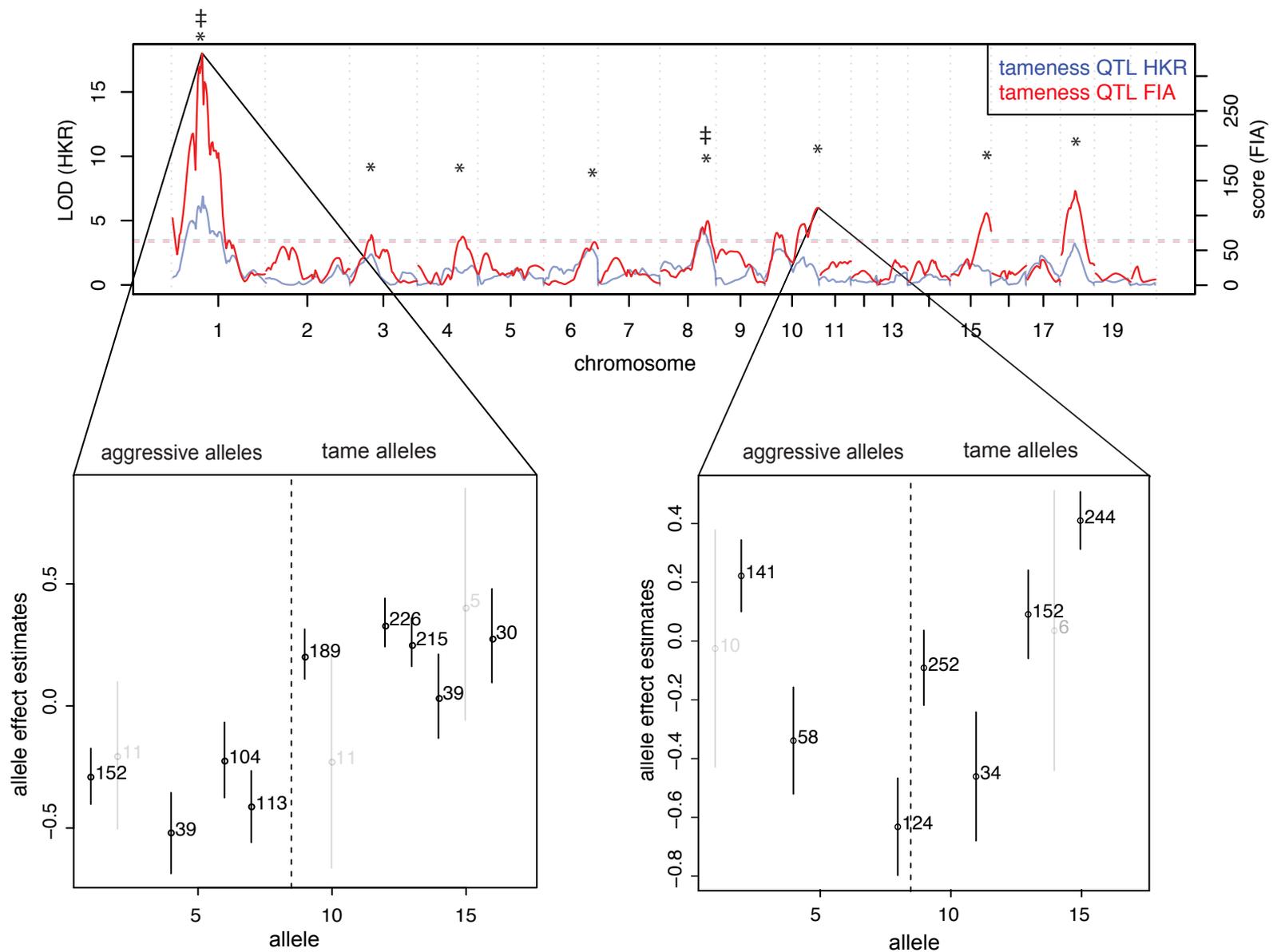

**Figure 1. Tameness QTL mapping.** Genome-wide statistical support for QTL affecting tameness in analyses assuming fixation of causal alleles within the founder populations (HKR – blue line) as well as analyses allowing segregation of causal alleles among founder animals (FIA – red line). Horizontal dashed lines indicate a 2% FDR significance threshold; QTL exceeding the threshold are indicated by * (FIA) and ‡ (HKR). Allele effect plots of the FIA tameness QTL on chromosomes 1 and 10 are shown below. Circles indicate the point estimate, lines the standard deviation. The numbers of individuals that carry a respective allele are printed to the right of the point estimates. For clarity, the results for alleles with less than 20 observations are shown in light grey.

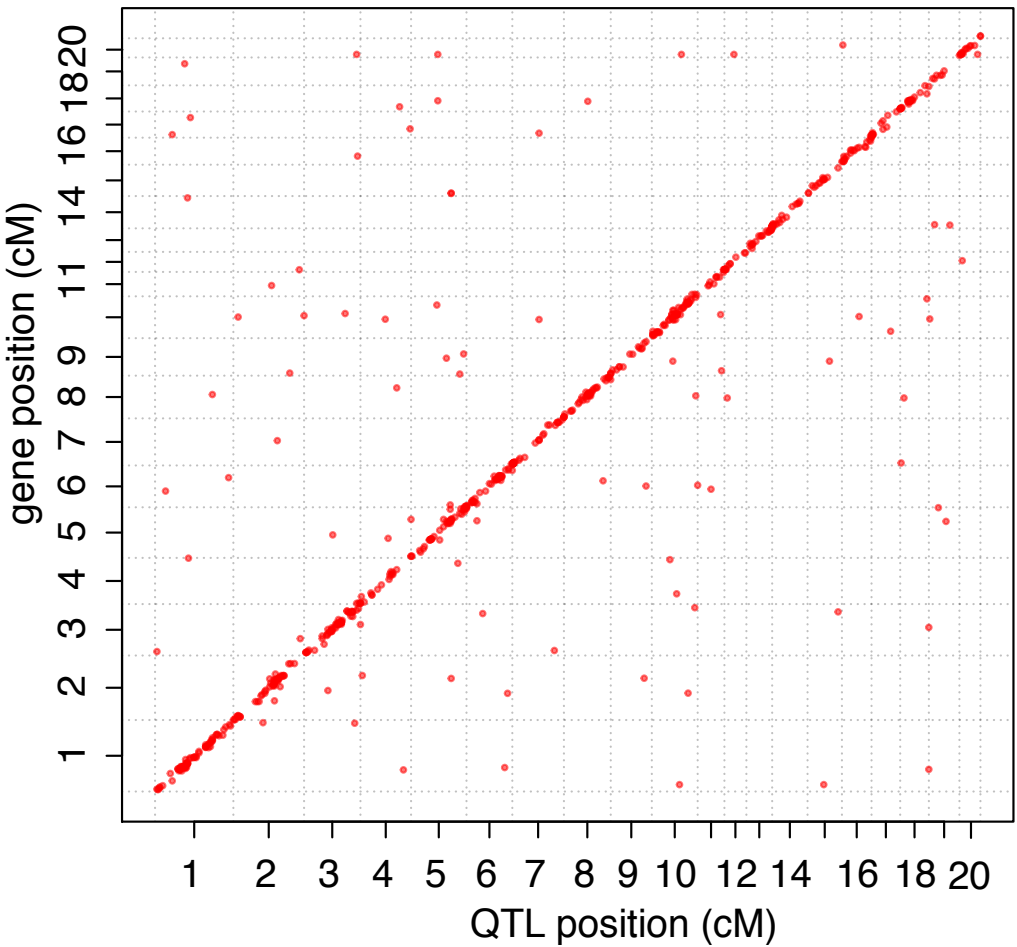

**Figure 2. Brain expression QTL mapping.** Locations of eQTL affecting brain gene expression in an F2 intercross between two rat lines selected for tame and aggressive behavior towards humans. Positions of genes are plotted against the positions of eQTL identified by FIA.

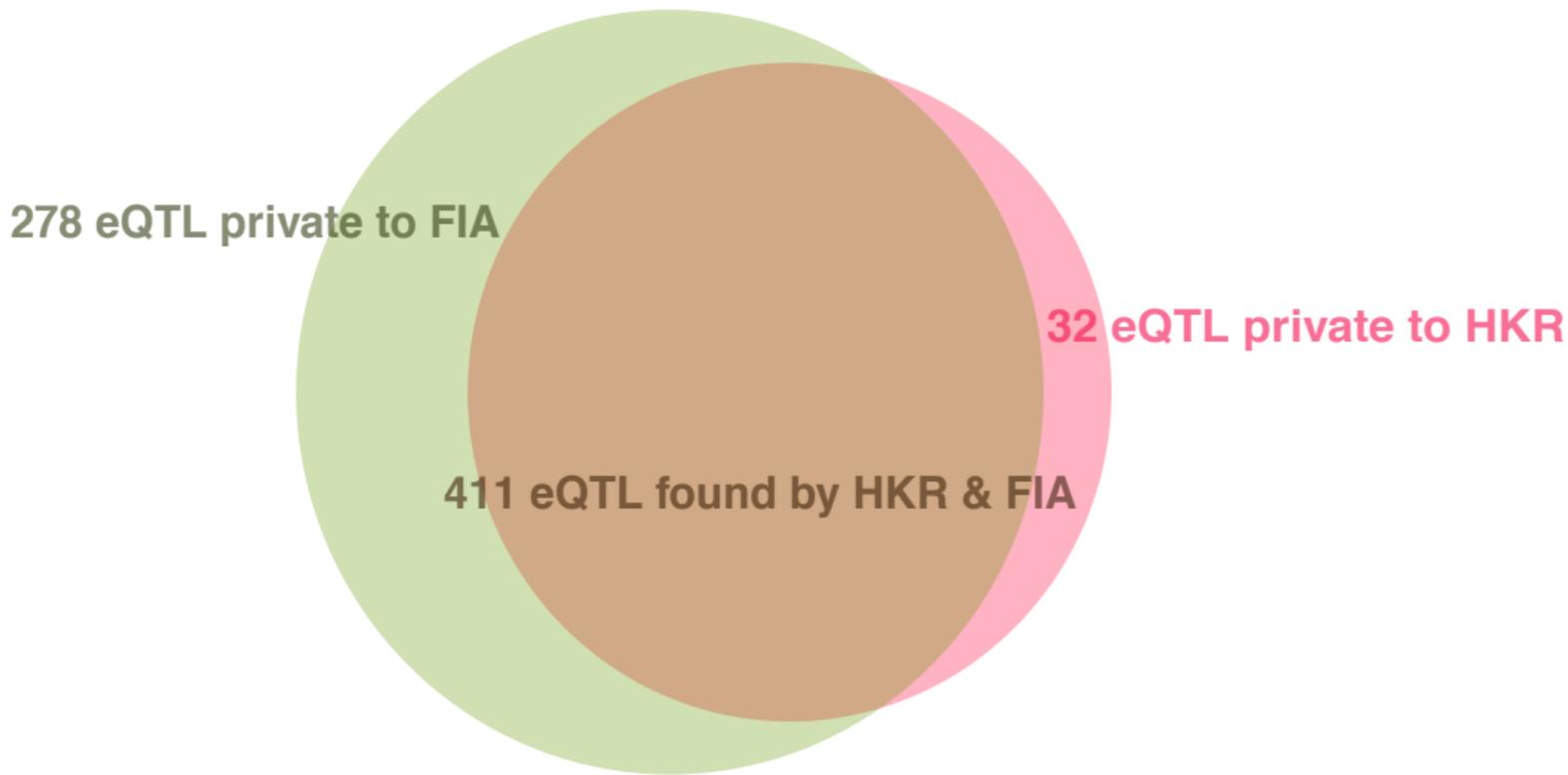

**Figure 3. Venn diagram of the number of eQTL identified by HKR and FIA.** Shown are eQTL identified at a 5% FDR.

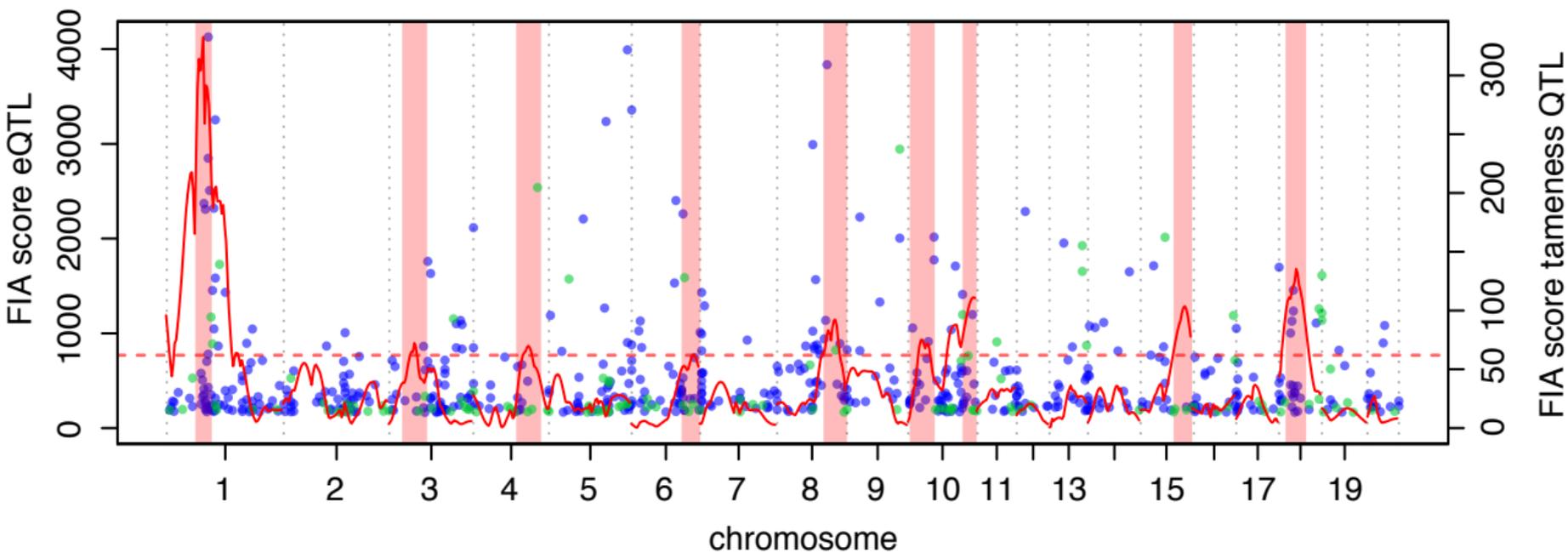

**Figure 4. Genome-wide overlap of QTL affecting tameness and eQTL affecting gene expression in the brain.** Red solid line: FIA scores for tameness QTL mapping. Light red rectangles: Confidence intervals for tameness QTL locations. FIA eQTL at a 5% FDR are shown as dots. Blue: local eQTL. Green: distant eQTL. Red dashed horizontal line: 2% FDR score threshold for tameness QTL.

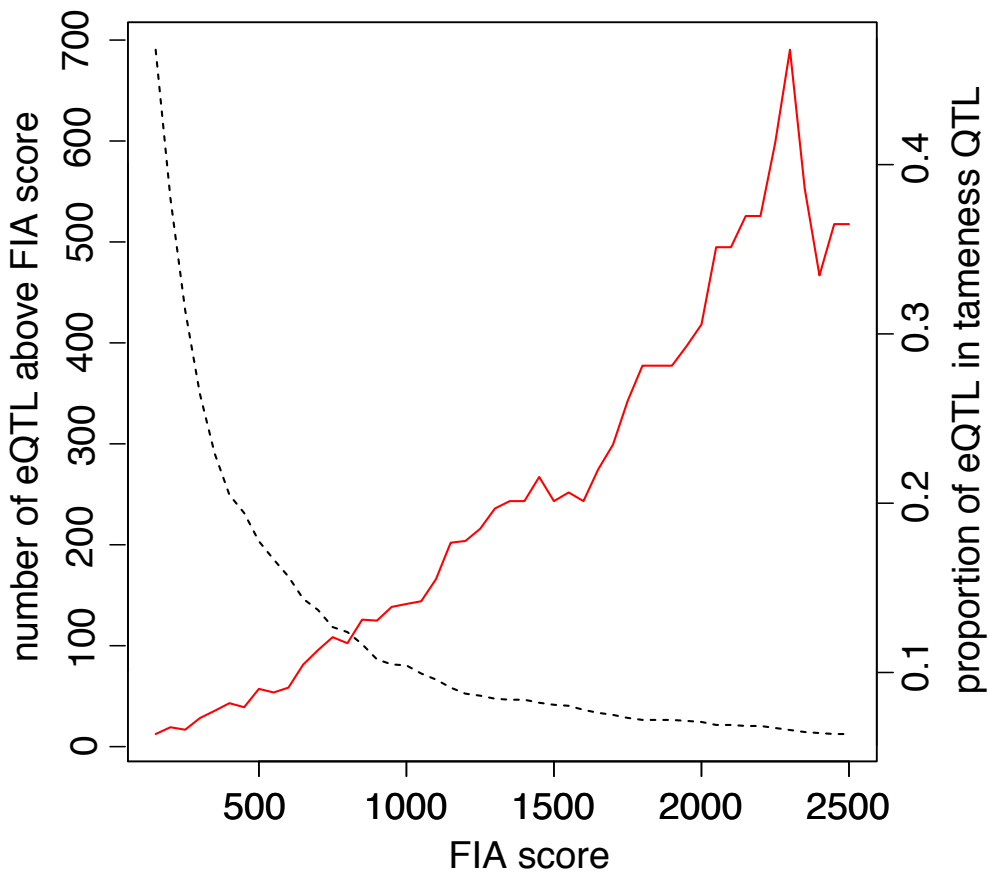

**Figure 5. Enrichment of highly significant eQTL at the Tame-1 locus.** The proportion of local and distant eQTL overlapping with Tame-1 (red line) at a given significance threshold (x-axis) is compared to the total number of all eQTL reaching the given level of significance (dashed black line).

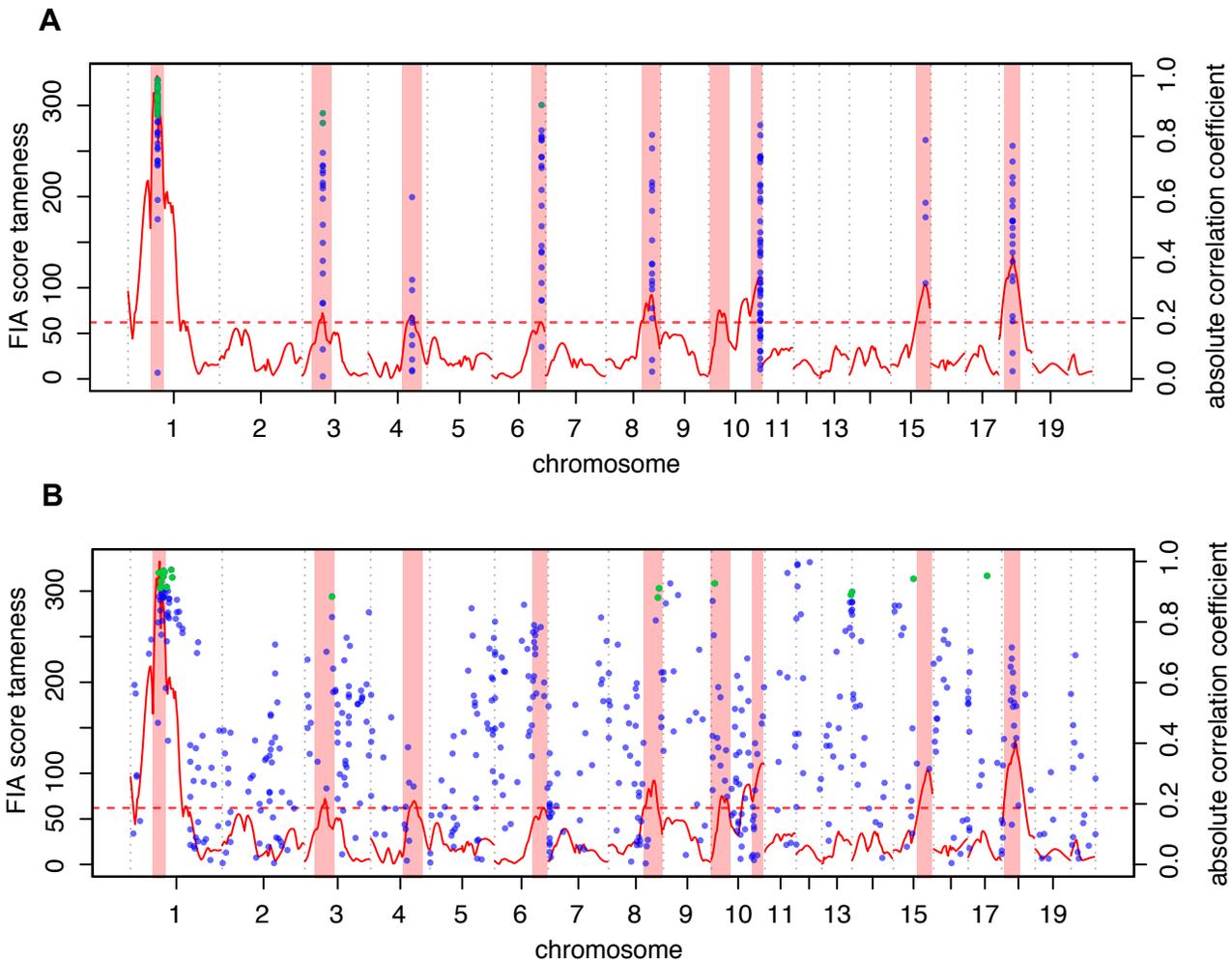

**Figure 6. Correlations between allele effects of eQTL and tameness QTL along the genome.** Green: significant (p < 0.05, Benjamini Hochberg corrected) correlations, blue: non-significant correlations. FIA mapping scores for tameness are shown as the solid red line and the confidence intervals for significant tameness QTL are shown as light red rectangles. Red dashed horizontal line: 2% FDR score threshold for tameness QTL. A. Local eQTL that overlap a tameness QTL tested at the peak of the corresponding tameness QTL. B. Allele effect correlations at all local eQTL peak positions. Note that 18 out of 23 significant allele effect correlations occur at local eQTL that overlap with a tameness QTL.

Zfp51, Siglec5,
Zfp719, D3ZT48_RAT,
Hddc2, Entpd5, Cyp7b1,
Clec11a, Prmt1, Pdk1,
Etfb, Gramd1a, Rtn2,
Fxyd1, Sirt2, Qk

Znf579, Snrpn,
Aldh16a1, Zscan18,
Mrps12, Phospho2,
Rps16,
Rps16's pseudogene
Zfp40

Dhdh, Slc17a7

Lgi4
Gltscr2

Mtmr15, Pccb,
Carhsp1, Plxnb1,
Esd, Eif3k

Dll3

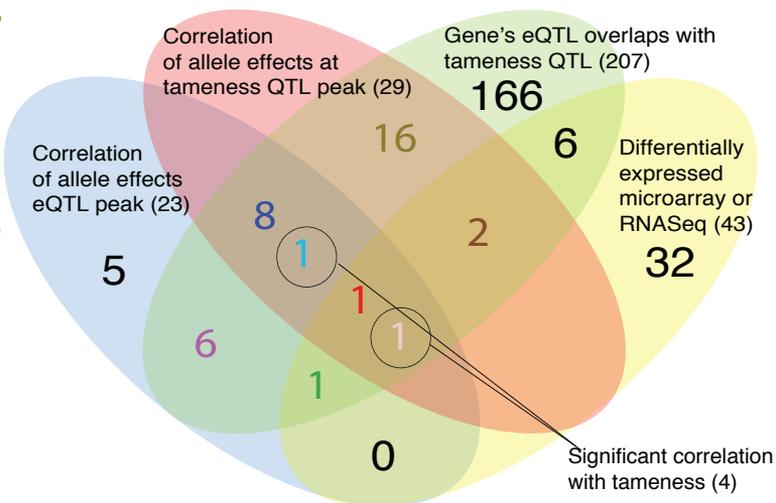

**Figure 7. Genes that show evidence for influencing tameness and aggression.** Shown are the number and names of genes that meet at least one of five criteria for involvement in tameness: i) positional overlap with tameness QTL (green oval); significant allele effect correlations with tameness QTL at either ii) eQTL peaks (blue oval) or iii) tameness QTL peaks (red oval); iv) differential expression in RNA-Seq frontal cortex or in whole brain microarray between the tame and aggressive populations (yellow oval); v) significant correlation with tameness in F2 animals (black circles). The numbers in parentheses indicate the number of genes that fulfilled the respective criterion. Gene names given on the left are color-coded correspond to those in the figure and are based on the criteria the given gene matches. Note that the overlap between the two scans for allele effect correlations is based on strict significance criteria in both scans (Methods), resulting in a conservative estimate of the intersection.

TABLES

*Table 1 – Locations of FIA tameness QTL*

| Chromosome | QTL start (cM) | QTL peak (cM) | QTL end (cM) | QTL start (Mb) | QTL peak (Mb) | QTL end (Mb) |
|---|---|---|---|---|---|---|
| 1 | 40.4 | 52.0 | 63.6 | 43.7 | 69.4 | 94.1 |
| 3 | 18.8 | 36.0 | 53.2 | 25.1 | 41.0 | 66.9 |
| 4 | 60.8 | 78.0 | 95.2 | 107.1 | 140.0 | 163.7 |
| 6 | 70.8 | 88.0 | 96.0 | 112.0 | 131.3 | 144.0 |
| 8 | 65.7 | 82.0 | 98.0 | 92.6 | 106.6 | 126.8 |
| 10 | 2.9 | 20.0 | 37.1 | 7.6 | 23.3 | 38.4 |
| 10 | 76.4 | 92.0 | 96.0 | 90.2 | 103.3 | 106.0 |
| 15 | 46.1 | 62.0 | 72.0 | 72.5 | 91.3 | 104.0 |
| 18 | 9.1 | 24.0 | 38.9 | 14.6 | 48.2 | 62.7 |

The locations of all FIA tameness QTL identified at a 2% FDR are shown.